\documentclass[12pt,article,floats,superscriptaddress]{revtex4}
\usepackage{graphicx}
\usepackage{dcolumn}
\usepackage{bm}
\usepackage{comment}
\usepackage{natbib}
\usepackage{wrapfig} 
\begin{document}
\title{Topological protection revealed by real-time longitudinal and transverse studies}

\author{Hoai Anh Ho}

\author{Jian Huang}
\email{jianhuang@wayne.edu}
\affiliation{%
Department of Physics and Astronomy, Wayne State University, Detroit 48201, Michigan, USA\\
}%
\author{L. N. Pfeiffer}
\author{K. W. West}
\affiliation{
Department of Electrical Engineering, Princeton University, Princeton 08544, New Jersey, USA \\
%
}


\date{\today}

\newpage
\begin{abstract}
\textbf{
Topology provides an essential concept for achieving unchanged (or protected) quantum properties in the presence of perturbations~\cite{TI_Rev,TI_SC_Rev}. 
A challenge facing realistic applications is that the level of protection displayed in real systems is subject to substantial variations. Some key differences stem from mechanisms influencing the reconstruction behaviors of extended dissipationless modes. Despite various insightful results on potential causes of backscattering, the edge-state-based approach is limited because the bulk states, as shown by breakdown tests, contribute indispensably. This study investigates the influence of bulk reconstruction where dissipationless modes are global objects instead of being restricted to the sample edge. An integer quantum Hall effect (IQHE)~\cite{klitzing1980new} hosted in a Corbino sample geometry~\cite{Halperin1982} is adopted and brought continuously to the verge of a breakdown. A detection technique is developed to include two independent setups capable of simultaneously capturing the onset of dissipation in both longitudinal and transverse directions. The real-time correspondence between orthogonal results confirms two facts. 1. Dissipationless charge modes undergo frequent reconstruction in response to electrochemical potential changes, causing dissipationless current paths to expand transversely into the bulk while preserving chirality. A breakdown only occurs when a backscattering emerges between reconfigured dissipationless current paths bridging opposite edge contacts. 2. Impurity screening is vital in enhancing protection, and topological protection is subject to an intriguing interplay of disorder, electron-electron interaction, and topology. The proposed reconstruction mechanism qualitatively explains the robustness variations, beneficial for developing means for optimization.
}
\end{abstract} 

\pacs{Valid PacS appear here}
\maketitle


\section{introduction}

A fundamental consequence of topology is the extended dissipationless states usually confined to system boundaries, free from backscattering~\cite{TI_SC_Rev}. On general grounds, the mechanism prohibiting backscattering derives from the symmetries, i.e. charility in quantum Hall effect~\cite{Halperin1982,Buttiker1988,hatsugai1993chern} and quantum anomalous Hall (QAH) effect~\cite{liu2016quantum}; time-reversal symmetry in the presence of spin-orbit coupling for quantum spin Hall (QSH)~\cite{kane2005z} and topological insulators (TIs)~\cite{TI_Rev,TI_SC_Rev}. However, when exposed to sufficient external excitation, vulnerability emerges as deviations from expected quantization, often called breakdowns. As demonstrated in the IQHE, deviations from quantized Hall conductance [$(1/\nu) e^2/h$] (in response to the sufficient external electric field) are tied to drastic changes in the transverse-direction properties and, thus, are bulk behaviors. The charge transport characteristics vary from percolation~\cite{Trugman1983,chalker1988percolation,lee1993quantum,furlan1998electronic} to dielectric-like behaviors~\cite{Cage1983BreakdownDef} in a highly system-dependent manner. $\nu$ is the filling factor, $e$ is the electron charge, and $h$ is the Planck constant. The robustness of protected states is reflected by the value of the threshold breakdown field and is a system-dependent variable even for the same filling factor~\cite{Cage1983BreakdownDef,haremski2020electrically,appugliese2022breakdown}. Meanwhile, deviations from Hall quantization hosted in systems such as graphene appear different because they occur readily, even at tiny biases. Though not usually categorized as breakdowns due to smaller scales, the precision level represents nearly three orders of magnitudes reduction~\cite{tzalenchuk2010towards} than in silicon and III–V heterostructure devices where up to parts per billion precision is observed~\cite{von2005developments}. The underlying mechanisms are fundamental to broad topological systems. The current study demonstrates that the discrepancies originate from a global mechanism, and the robustness of the dissipationless modes depends on the bulk localization effect influenced by disorder, electron-electron interaction, and topology. 

Application of an external electric field~\cite{haremski2020electrically} effectively tunes the ``distance" to the breakdown point at a controllable precision. This is beneficial for detecting how the extended dissipationless modes reconfigure dynamically, a key concept that is only explored in a limited sense by edges-state studies~\cite{chamon1994sharp,yang2003field,bid2010observation} since the bulk states (or rather the bulk degrees of freedom) are not involved. A viable means is to measure the breakdown transport characteristics longitudinally and transversely. The two directions are drastically anisotropic as one is sustained by chiral currents and the other by bulk localization. Such experiments are usually nontrivial for several reasons. 

The bulk states are complicated due to spatially varying energy profiles due to random disorder fluctuations. In the disorder-dominated limit, the bulk states are described by the one-particle percolation picture~\cite{Trugman1983,chalker1988percolation,lee1993quantum,furlan1998electronic}. For a fluctuation wavelength $\lambda$ far exceeding the magnetic length $l_B=\sqrt{\hbar/eB}$, charges residing on the contours of the disordered potential become extended only when the Fermi level falls in the mobility edge. The 2D areal space divides into local regions where the Landau level (LLs) are either full or empty~\cite{Trugman1983}. $B$ is the perpendicular magnetic field. This picture, however, is qualitatively altered if electron-electron interaction is important. A vital effect is the screening of charged impurities which rises with the emergence of a third kind of region of metallic nature between filled and empty regions~\cite{cooper1993coulomb}. Screening of long-range impurities is expected to drastically alter bulk localization because the charges residing on different contours of the disordered potentials can now differ significantly in the potential energy~\cite{haremski2020electrically,tsemekhman1997theory}. The sudden breakdown behaviors~\cite{cooper1993coulomb,tsemekhman1997theory} observed previously are believed to be relevant, even though the dissipation mechanism is not adequately detectable with conventional setups. 

\begin{figure}[b]
\vspace{0pt}
\includegraphics[width=0.8\textwidth]{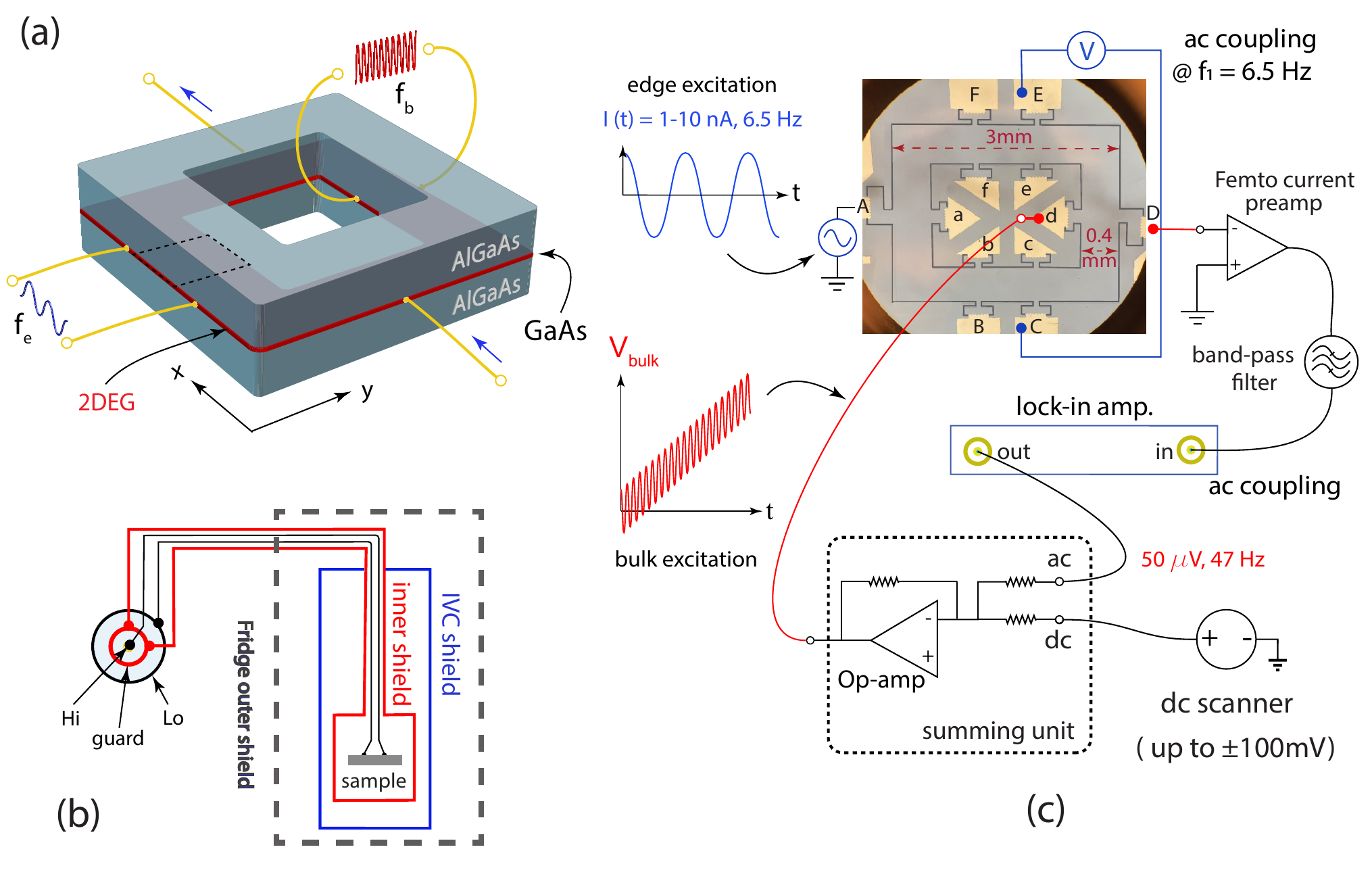} 
\vspace{-15pt}
\caption{\label{fig:Fig1} (a) shows schematically an anti-Hall bar quantum well system under simultaneous testing of dissipation current(s) along the longitudinal (x) and transverse (y) directions. (b) is the schematics for the guarded measurement setup. (c) shows a photo of the Corbino device with twelve Ohmic contacts. The dual measurement setup consists of two independent sources, marked in red and blue, for simultaneous detection in orthogonal directions. A diagram including a dc$+$ac sourcing and a current preamp for sensing are also shown.}
\end{figure}

The system chosen for our study is the IQHE hosted in high-quality two-dimensional systems in GaAs quantum well structures, known to be robust and representative. Other topological orders, such as QAH, QSH, and TI, are also expected to work. The strategy is to utilize a rectangle Corbinal (anti-Hall bar) geometry~\cite{Laughlin1981,Halperin1982,SyphersCorbino1986,FonteinCorbino1988,ebert1985hall} as schematically shown in Fig.~\ref{fig:Fig1}(a), which enables independent sourcing and sensing in the transverse (y-) direction with inner and outer contacts; while in the longitudinal (-x) direction, with contacts on the same edge. A variable dc voltage is maintained between the inner and outer edges, introducing an in-plane electric field (\textbf{E}) to drive it towards the breakdown. Two electric signals from different frequency bands ($f_e$ and $f_b$ marked in blue and red) are adopted for sourcing in orthogonal directions. Operating under ac coupling mode, the phase-locking detection in the transverse direction captures responses only to $f_b$ signal excitation, excluding $f_e$ contribution. Similarly, the edge detection captures responses to the $f_e$ signal excitation, excluding the $f_b$ contribution. This method allows the adoption of an independent electrometer setup suitable for sensing tiny changes in a highly insulating regime. A brief discussion is provided in Supplementary Materials B.

The results are reported in two steps. First, transverse detection of dissipation is conducted in real-time with the Hall measurement, confirming the bulk dissipation events responsible for the breakdown. Second, the same transverse detection is performed in real-time with longitudinal dissipation. Global reconstruction is then verified by comparing the detected responses in the time domain through whether the measured dissipation in orthogonal directions shares the same origin. The reconstruction mechanism is analyzed according to the onset characteristics of the breakdown current. A qualitative reconstruction model is presented by incorporating previous findings on bulk reconstruction in the presence of electron-electron interaction and scattering effects on chiral currents.

\section{Results}
The schematics of the first measurement are shown in Fig.~\ref{fig:Fig1}(c) consisting of a 2mm x 3mm anti-Hall bar device and a circuit diagram for two independent arrangements. The Hall voltage is measured with leads E and C in response to an ac drive of 1-10 nA at 6.5 Hz through leads A and D. For bulk detection, dc+ac sourcing is adopted between inner and outer electrodes (leads d and D), with the ac component confined to a  different frequency band (51-67 Hz). The dc component produces an in-plane transverse \textbf{E}-field ($V_{dc}$) orthogonal to the sample edge. It slowly varies the \textbf{E}-field when approaching the breakdown point at a precise step of 10$\mu e$V. Mixing the ac and dc components is realized through an ultra-low noise signal summing device. Bulk sensing is challenging due to the extraordinary bulk resistance ($R_b$) beyond G$\Omega$ in the studied cases. Large $R_b$ is notorious for debilitating typical transport and spectroscopic techniques such as electron compressibility and quantum capacitance measurements (due to squeezed bandwidths). The following design achieves sub-$p$A sensing resolutions under such extreme conditions. An ultra-low noise current preamp, Femto LCA-200k-20M, is adopted for the sensing. The wiring connection is a triaxial setup adapted to implement a guarded enclosure [see 
Fig.~\ref{fig:Fig1}(b)]. It isolates the high impedance input and, more importantly, cuts down the response delay (which easily reaches up to 60 seconds for a typical 100 $n$F stray capacitance if without the guard). Noise filters are also implemented to minimize signal distortion. The option of using voltage sensing (with current sourcing) is also tested with a suitable electrometer setup with $10^{10}\Omega$ input impedance, and the information is provided in Supplementary Materials B. The sample characterization (including the decoupling of the inner and outer edges) is provided in Supplement Materials A.  

\begin{figure}[b]
\includegraphics[width=0.9\textwidth]{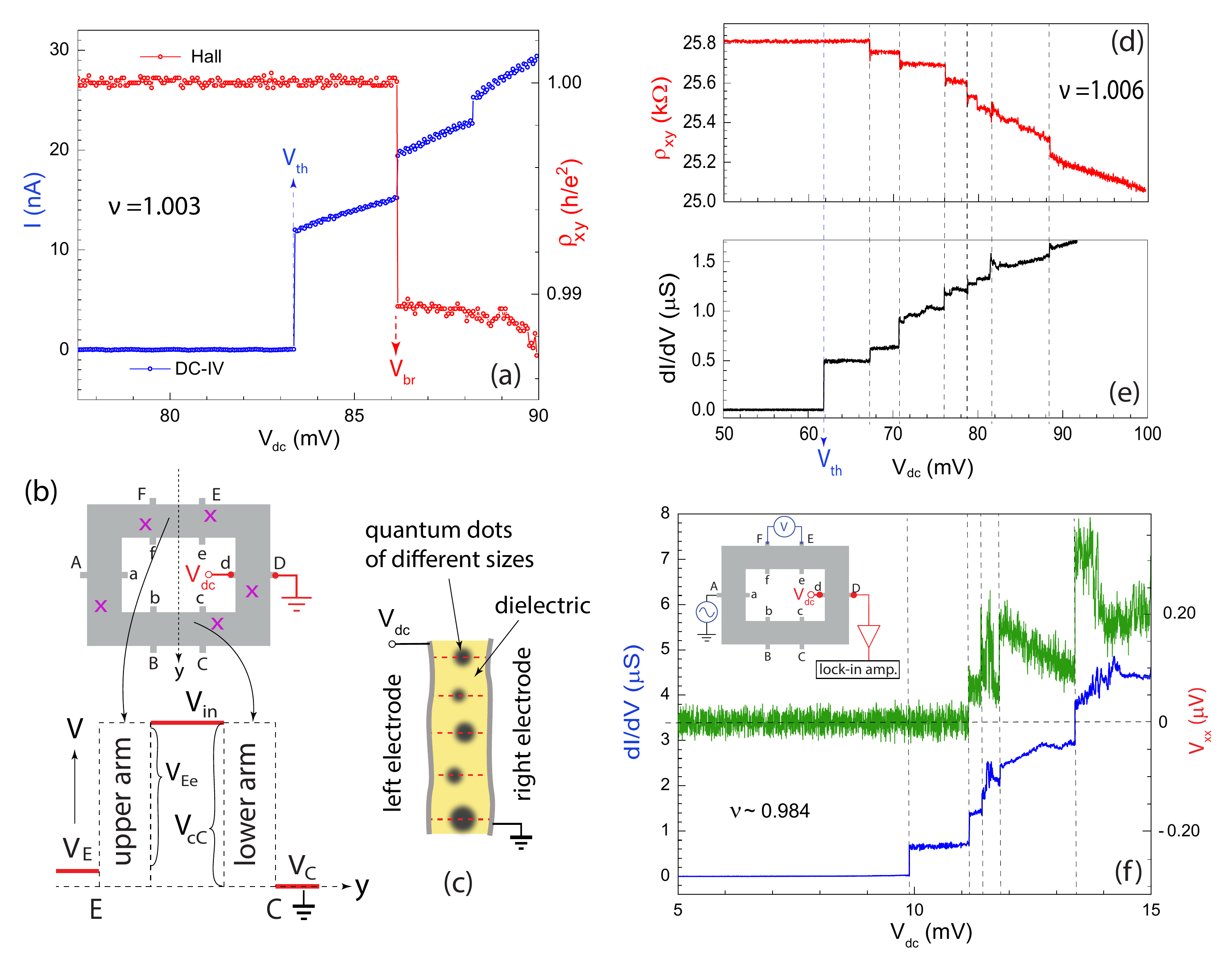}
\vspace{-10pt}
\caption{\label{fig:Fig2} The bulk dc-VI at $\nu=1.003$ is shown in (a) in comparison to the real-time $\rho_{xy}$. The coincidence starting from the second jump is because, along a cross-section (dotted) line through the anti-Hall bar, as shown in (b), there is a finite electrical potential imbalance between the upper and lower arms near a breakdown. The ``x" labels in the anti-Hall bar represent random tunneling sites causing the ``staircase" dc-VI response. Resonant tunnelings produce the same behavior through a parallel tunneling network of quantum dots sandwiched between parallel electrodes shown in (c). (d) and (e) show the real-time correspondence between $\rho_{xy}$ and the bulk $dI/dV$ measured at $\nu=1.006$ with the setup shown in Fig.~\ref{fig:Fig1}(c). The real-time correspondence between bulk $dI/dV$ and the edge potential change is captured in (f).}
\end{figure}

First, bulk dissipation is detected with a dc-VI, with the ac component (shown Fig.~\ref{fig:Fig1}(c)) kept at zero. The result shown in Fig.~\ref{fig:Fig2}(a) is obtained for $\nu\sim1.003$ along with the Hall resistance $\rho_{xy}$ measured simultaneously. In contrast to percolation, a bulk current emerges as a sharp threshold jump, from zero to nearly 12 nA, at 83.5 mV (which is referred to as $V_{th}$ hereafter). No breakdown emerges at $V_{th}$ since quantized $\rho_{xy}$ at $h/e^2$  remains unperturbed. However, at $V_{dc}=86$ mV where a second jump occurs, $\rho_{xy}$ exhibits a sharp downward jump simultaneously by roughly 1\%. The corresponding $V_{dc}$ is referred to as $V_{br}$ to mark the onset of a breakdown.

The breakdown starting only from the second jump is consistently detected in all the dc-VI performed at various filling factors. This behavior is caused by the imbalance of biases between the upper and lower Corbino arms due to two factors. First, the edge current excitation introduces a finite Hall voltage between the upper and lower outer edges. Along the dotted cross-section line shown in Fig.~\ref{fig:Fig2}(b), there is always a time-dependent imbalance of biases from the inner edge to the upper and the lower outer edges. Second, the differences in the disorder levels in the two arms also alter the breakdown threshold (though likely a smaller effect). Consequently, the arm under a larger bias always breaks down first. The second jump in the dc-VI signifies a bulk current running across the other arm, which causes a breakdown as indicated by deviation in $\rho_{xy}$ from Hall quantization. Thus, global dissipationless modes remain protected as long as the two outer edges remain isolated. The changes in $\rho_{xy}$ always decrease, indicating that the Hall resistance is the upper limit of a permissible $R_b$. 

\begin{figure}[h]
  \centering
  \includegraphics[width=0.8\textwidth]{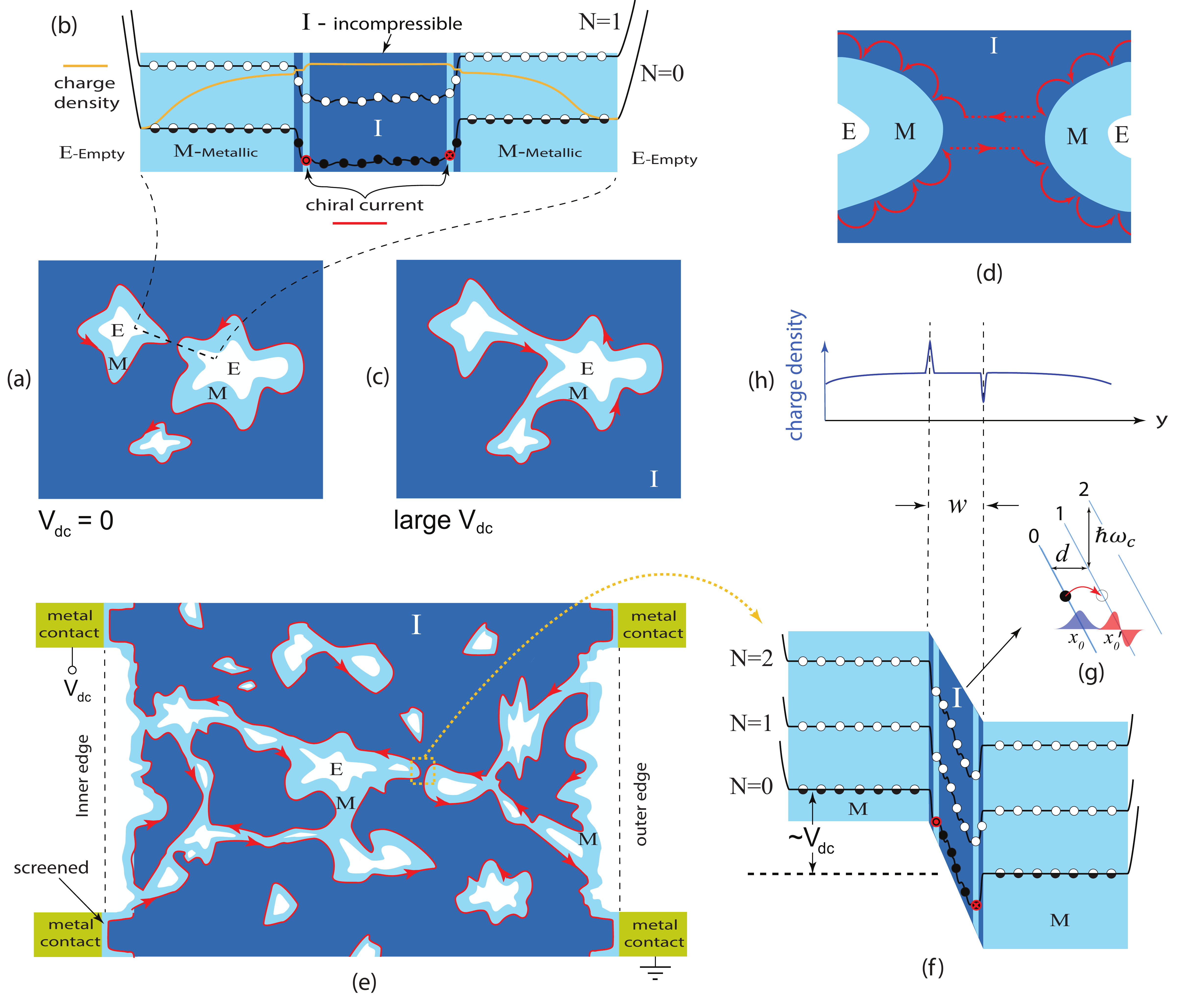}
  \vspace{-10pt}
  \caption{(a) shows a small bulk area containing three disorder sites for $V_{dc}=0$, with the incompressible regions (dark blue), metallic regions (light blue), and empty regions (white). The chiral current loops surrounding the metallic region are marked in red. (b) shows, for $\nu\sim 1$, the changes in the first two LLs along a line segment from one empty region to another. The merger of the regions is shown in (c) at increased $V_{dc}$. (d) illustrates the resonant tunneling between two skipping orbits. (e) illustrates the situation on the verge of a breakdown when the last incompressible strip separates two globally reconstructed dissipationless current paths. The condition for inter-LL resonances is illustrated in (f) and (g) when the wavefunctions of the first and second LLs overlap. (h) is the corresponding charge distribution showing a dipolar charge accumulation at the strip boundaries.
}
  \label{fig:Fig3}
  \vspace{10pt}
\end{figure}

The ``staircases" dc-VI characteristic is almost identical to the resonant tunneling transport through some parallel silicon quantum dots embedded inside a dielectric media sandwiched between parallel electrodes~\cite{boeringer1995avalanche}. As illustrated in Fig.~\ref{fig:Fig2}(d), the inhomogeneity of the quantum dots causes the tunneling to be turned on one dot at a time by increasing bias. Here, the same multi-threshold behavior can be understood as tunnelings through some random sites inside the antiHall bar arms as illustrated in Fig.~\ref{fig:Fig3}(b). The staircase provides important clues to a reconstruction distinctive from the non-interacting percolation transition or (thermally) activated transport~\cite{Trugman1983,chalker1988percolation,lee1993quantum,furlan1998electronic}, in which inhomogeneous broadening due to a globally distributed large number of hopping sites smears individual tunneling peaks. 

A more sensitive probe to the onset of bulk dissipation is then adopted for detecting dynamic changes. The ac component shown in Fig.~\ref{fig:Fig1}(c) is turned on to $\sim50\mu$V oscillation at $67$~Hz, turning it into a differential conductance ($dI/dV$) setup more sensitive to dynamic changes. Note that the transverse signal response is band-pass-filtered before the lock-in input, eliminating potential mixing from the ac edge excitation. Fig.~\ref{fig:Fig2}(d) and (e) show the real-time $dI/dV$ and the Hall results for $\nu=$1.006. An extremely sharp jump in bulk current occurs at $V_{th}=$62~mV. The decrease in $V_{th}$ is substantial considering a slight change of $\nu$. Indeed, $V_{th}$ exhibits exponential dependence on $nu$. Starting from the second jump at $V_{br}=65.5$mV, the staircases correspond perfectly to every jump. This is consistent with the dc-VI result. The number of jumps increases noticeably, indicating more tunneling sites. The $\rho_{xy}$ steps show weak dependence on $V_{dc}$ first and become progressively less prominent at higher $V_{dc}$ before eventually diminishing around $V_{dc}\sim90$mV, beyond which $dI/dV$ resembles an avalanche effect in dielectric breakdowns. The $\rho_{xy}$ steps mark a 1-2\% decrease in the unit of $h/e^2$. Two observations support resonant tunneling. One, the threshold has a line width $\leq$ 10$\mu e$V (the limit of the scan resolution), consistent with a single energy behavior. In another differential conductance ($dV/dI$) measurement, the jump represents almost four orders of magnitude change, from $>$50~G$\Omega$ to 8~M$\Omega$, over a single step of 10~$\mu$V increase in $V$ [see Fig.~\ref{fig:appen1}(e) in Supplementary Materials]. The corresponding field strength, as shown later, fits well with the expectation for inter-LL resonances. Two, the bulk remains a strong insulator marked by ~$\sim$8-10M$\Omega$ resistance (or resistivity of $\sim$100 M$\Omega$ per square) far exceeding the quantum resistance ($h/e^2$), excluding the presence of metallic paths. 

To verify the relationship between the transverse bulk response and the longitudinal response, the Hall measurement [in Fig.~\ref{fig:Fig1}(c)] is replaced with a magnetoresistance setup (with electrodes on the same sample edge) [inset of Fig.~\ref{fig:Fig2}(g)]. Detection of longitudinal dissipation is achieved with sensing deviations in $V_{xx}$ from zero, and it is measured in real-time with transverse dissipation at 20~$n$V sensing resolution. The ``staircase" bulk dissipation provides unique markers for comparing the longitudinal dissipation behavior. Here, the filling factor is randomly chosen to be $\nu=0.984$. The first jump in $dI/dV$ occurs $V_{th}$ at 10 mV where zero $V_{xx}$ is unperturbed. However, at the second jump at $V_{dc}\sim11.2$mV, $V_{xx}$ jumps sharply from zero to approximately 60~nV. Again, the one-to-one correspondence happens at each jump in the same manner shown in Fig.~\ref{fig:Fig2}. These results confirm a perfect correspondence between the longitudinal and transverse behaviors, affirming rare-site resonant tunnelings as the common origin for altering the measured topology in orthogonal directions (Hall and magnetoresistance). 

For backscattering to occur between dissipationless current paths connected to opposite edges, the dissipationless current paths must be rerouted into the bulk at large \textbf{E}-fields. The proposed model highlights a mechanism preserving chirality. The key ideas are illustrated by some out-of-scale cartoon drawings in Fig.~\ref{fig:Fig3} where the screening of long-range impurity plays an essential role in influencing the reconstruction process~\cite{cooper1993coulomb,tsemekhman1997theory}. The importance of electron-electron interaction is reflected in two aspects. First, the Coulomb energy $E_c$ exceeds the cyclotron energy $\hbar\omega_c$ by approximately sixty times near $\nu=1$. The exceptional charge mobility found in the zero B-field confirms the low disorder nature, which is essential for not overwhelming the interaction effect through disorder localization~\cite{knighton2018evidence}. Second, to have a dielectric breakdown at enormous threshold fields, resonant tunneling must happen between separated chiral paths held at significantly different potentials, which is not permissible in the one-electron model. Instead, it fits the characteristic of a self-consistent potential resulting from the screening effect in response to the induced electrochemical potential gradient. Previous compressibility studies probed this behavior by tracing the potential variation as a function of charge density and the B-field~\cite{Yacobi2004microscopic}. 

At small or zero $V_{dc}$, incompressible regions with complete filling percolate throughout the 2D areal space for $\nu$ close to the integer. What influences the breakdown response is the disorder sites with strong potential fluctuations where the screening is imperfect because the required charge density modulation exceeds the finite charge density in the LL. Consequently, as illustrated in Fig.~\ref{fig:Fig3}(a) for a tiny area containing three disorder sites, ring-shaped metallic regions (marked in light blue) spread around the strong disorder sites (marked in white) due to Coulomb repulsion~\cite{efros1988non,cooper1993coulomb}. The chiral currents form around the disorder potential sites in an identical manner as at the sample edge, except they are isolated closed loops. As presented in Ref.~\cite{Electrostatics_of_edge_channels}, the separation of the 1D chiral current from the metallic region by an incompressible strip is due to filling-factor-dependent charge screening. This screening differs from the impurity screening, which depends primarily on the gradient of disorder potential fluctuation instead of $\nu$. The local electric field across the incompressible strips is roughly $\Delta V_{dis}/w$ where $\Delta V_{dis}$ is the disorder potential difference, not yet sufficient to cause mergers of the metallic regions. This is schematically shown in Fig.~\ref{fig:Fig3}(b) with the lowest two LLs (under the influence of disorder fluctuation) along a line segment crossing the disorder site. The average incompressible width $w$ separating neighboring metallic regions is on the order of $\lambda$ much greater than $l_B$.

Increasing $V_{dc}$ raises the \textbf{E}-field across the incompressible regions. As previously shown~\cite{Tsemekhman1997}, mergers start at the narrowest incompressible strips when the local \textbf{E}-field is on the order of $\sim\hbar\omega_c/e l_B$. With an expelled electric field, the former incompressible region is now effectively a part of a larger compressible region held at equipotential [see Fig.~\ref{fig:Fig3}(c)]. The \textbf{E}-field then falls on the remaining incompressible strips. The merging mechanism needs to satisfy two conditions based on the results -  One, it involves resonant tunneling; Two, zero longitudinal $V{xx}$ is unaffected by the mergers, meaning the absence of backscattering until the breakdown starts. 

The proposed merger via resonant tunneling between two chiral paths is depicted in Fig.~\ref{fig:Fig3}(d). When a charge in a skipping orbit around a chiral path tunnels to the other, it joins the skipping orbit within a relaxation length on the order of a cyclotron radius. If the destination is a closed-loop chiral current path, tunneling leads to two consequences. One, the chemical potentials of the two chiral paths eventually equilibrate when the tunneling rates in both opposite directions match. The potentials of the metallic regions settle self-consistently. Two, chirality is preserved on the reconstructed dissipationless path, which is now on a greater length scale. Such reconstruction naturally maintains zero longitudinal $V{xx}$. This is because as frequent mergers expand the chiral current paths deeply into the bulk with rising \textbf{E}-field, the electrodes on the same edge remain connected through chiral currents. For a small section of the Hall bar arm in which the metallic networks expand into the bulk, Fig.~\ref{fig:Fig3}(d) illustrates the upper limit for a zero $V_{xx}$ where two significantly reconstructed dissipationless current paths, each connected to the contacts on opposite edges, are separated by the last incompressible strip. The drawing shows an unrealistic small amount of disorders for a visual illustration. Unlike for the closed loops, tunneling across this last barrier results in dissipation and breakdown. 

The single-energy resonances are characteristic to inter-LL resonance~\cite{QUILL} across a finite width of an incompressible strip shown in Fig.~\ref{fig:Fig3}(e) containing a large number of tilted LLs~\cite{QUILL}. Fig.~\ref{fig:Fig3}(f) shows schematically the band diagram for the three lowest tilted LLs under bias across the compressible width $w$. Increasing the \textbf{E}-filed ($\approx V_{dc}/w$) shrinks the horizontal spatial distance ($d$) between neighboring LLs towards $l_B$. A resonant condition in the spinless limit is when $d\geq S_{N=0}(x_0)+S_{N=1}(x'_0)$ approaches $l_B$. $S_N=\sqrt{<\Delta x^2>}=(2N+1)^{1/2}l_B$ is the spatial spread of the wavefunction for Nth LL. Fig.~\ref{fig:Fig2}(f) illustrates the resonances between the filled N=0 and empty N=1 LLs when the critical field is approximately $E_{br}=\hbar\omega_c/[(1+\sqrt{3})el_B]\sim10mV/\mu$m. As a result, charges tunnel through the barrier by occupying higher LLs. For $V_{br}\sim70$~meV, the estimated number (N) of LLs corresponding to $V_{br}$ is $eV_{br}/\hbar\omega_c$ which is $\sim140$, or a $w$ of approximately 7.1~$\mu$m. The associated threshold field, $10me$V$/\mu$m (or 100V/cm), is enormous. Based on the second derivative of the tilted LLs, the charge density distribution shows a dipolar strip [shown in ]Fig.~\ref{fig:Fig2}(f)]. Note that $w$ depends sensitively on $\nu$ - the ratio of the compressible/incompressible areas increases as $\nu$ deviates further from the integer, resulting in smaller $w$. Resonant condition is then achieved at lower $V_{br}$. The result shown in Fig.~\ref{fig:appen1}(d) in Supplementary Materials indicates this dependence is exponentially sensitive. However, Fig.~\ref{fig:appen1}(f) shows the dielectric threshold behavior remains unchanged, indicating the same tunneling mechanism holds for all $\nu$ within the plateau. This feature supports impurity screening which only depends on the gradient of disorder potentials instead of $\nu$. 



\section{Summary Remarks}

Buttiker presents the original picture of preserving chiral currents through electron relaxation consequential to impurity scattering. Even though the actual chiral modes formed along equipotential edge contours can be quite complex~\cite{sabo2017edge,image2}, the basic idea is supported by the observed scattering effects~\cite{Frederic_EdgeSpectroscopy} as long as the current channels are well separated and equilibrated. 
The differences in the chiral modes can be qualitatively related to the charge screening effect, which depends sensitively on the local filling factor and the nature and concentration of the disorder potentials. These previous studies, nevertheless, unanimously concentrate on longitudinal effects that do not apply to bulk reconstruction. The proposed mechanism preserves chiral current in the transverse direction by connecting isolated chiral current loops (distributed around random disorder potentials) against the localization effect. The proposed mechanism is valid for the regime of screened long-range impurities. The reconstructed bulk state presents a different scenario than the seminal TKNN model~\cite{TKNN} (involving the entire system) since quantization is preserved as long as the opposite edge contacts are isolated, even by a single incompressible strip. It should be pointed out that previous breakdown models treat the mergers as tunneling between the metallic regions instead of between the chiral currents. A distinction is that tunneling between metallic regions necessarily results in deviated Hall conductance and edge dissipation, inconsistent with our results within the breakdown limit. In contrast, resonant tunneling across dipolar strips requires a significantly larger \textbf{E}-field on the order of $\sim\hbar\omega_c/e l_B$. This is supported by the exceptional breakdown threshold \textbf{E}-field up to $\sim10mV/\mu$m while the bulk remains insulating. Thus, protection is enhanced by electron-electron interaction despite being deemed unnecessary by the semiclassical description of the IQHE.

For systems falling between strongly interacting and non-interacting limits, a crossover from percolation to dielectric behaviors is anticipated as the ratio of screening length and disorder fluctuation wavelength $\lambda$ varies. Thus, changes in the disorders alter the breakdown behaviors for the same filling factor, resulting in the kind of discrepancies mentioned in the beginning. Also, breakdown behavior changes are expected in graphene systems where short-range impurities are abundant amid long-range impurities. Screening there depends on the effective dielectric constants~\cite{ando2006screening,kotov2012electron} and varies with the concentration of the disorder potentials and the charge density. Experiments indicated that charge delocalization is more prominent as indicated by a significantly smaller bulk resistance $R_b$ on hundreds of k$\Omega$~\cite{Yacobi2004microscopic,zibrov2017tunable}. Interestingly, this $R_b$ is approximately three to four orders smaller than typical IQHE bulk resistance in GaAs systems, echoing the quantum precision differences between the two systems. It is thus perceivable that, even in response to a small \textbf{E}-field, a fraction of the activated carriers percolate into the bulk and backscatter with the chiral charges connected to the opposite sample edges, resulting in deviations from the Hall quantization~\cite{aharon2021long}. Nevertheless, enhanced screening can still be found in such systems near electric contacts where an abundance of electrons are pinned at the Fermi level. This is why improved QHE is observed in edge-less graphene due to suppressed disorder fluctuations along the sample edge~\cite{polshyn2018quantitative,zeng2019high}. 

When discussing the inter-LL resonances, the effect exchange interaction, which splits the LLs into subbands, is ignored~\cite{usher1990observation,piazza1999first,nomura2006quantum} since it does not qualitatively alter the tunneling picture except for a slightly modified resonant condition. On the other hand, our results raise important perspectives on strongly correlated topological systems. First, the strong local many-body effect did not drive the fractional quantum Hall effect (FQHE). Understanding the interplay of electron-electron interaction and topology~\cite{hohenadler2013correlation,rachel2018interacting} requires further experimental demonstration elucidating the distinction from fractionalized many-body topological invariant~\cite{gavensky2023connecting}. Second, the spin-orbit coupling can be substantially enhanced by strong electron-electron interaction~\cite{huang2017spin}. This implies enriched quantum behaviors for cases of QSH and TI, which remain to be explored. Finally, on the experimental technique side, the reconstruction extends the reach of the edge electrode into the bulk through reconfigurable 1D dissipationless current paths, which presents a unique opportunity for employing various techniques, including spectroscopy, to probe bulk energy structures.

\section{acknowledgment}
The NSF supported this work under No.~DMR-1410302. The authors thank Professor Bertrand Halperin and Professor Joseph Avron for stimulating comments and discussions and Professor Klaus von Klitzing for raising important questions. The GaAs crystal growth work was partially funded by the Gordon and Betty Moore Foundation through Grant No.~GBMF2719 and by the National Science Foundation No.~MRSEC-DMR-0819860 at the Princeton Center for Complex Materials.

\newpage
\bibliographystyle{apsrev}

\bibliography{biblio.bbl}

\newpage
\section{Supplementary Materials}

\noindent\textbf{A. Sample and Characterization}

\begin{figure}[t]
\vspace{-10pt}\centering
\includegraphics[height=6.in,trim=0.0in 0.2in 0.0in 0.0in]{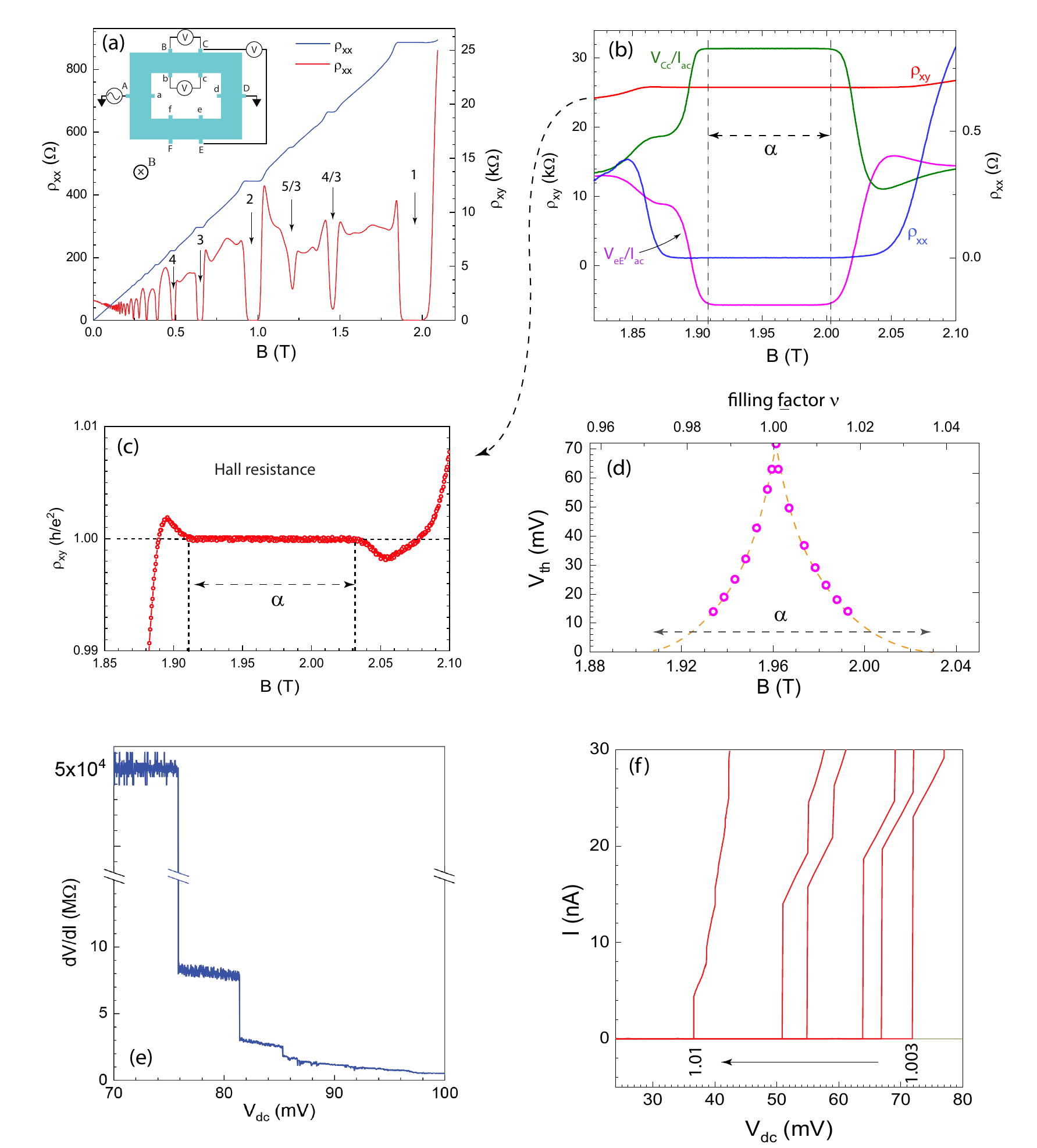}
\vspace{-5pt}
\caption{\label{fig:appen1} (a) Hall and magnetoresistance up to filling factor 1. (B) Voltage drops across upper and lower arms of the Corbino anti-Hall bar compared to the Hall and magnetoresistance near $\nu=1$. $\alpha$ marks the width for constant $V_{Cc}$ and $V_{eE}$. (c) shows the magnified $\rho_{xy}$ having the identical plateau width $\alpha$. (d) $V_{th}$ as a function of $\nu$. (e) Differential bulk resistance results for $\nu$ close to 1. (f) $\nu$ dependence of dc-VI curves showing reduced \textbf{E}$_{br}$ for $\nu$ from 1.003 (first on the right) to 1.01 (first on the left).}
\vspace{-0pt}
\end{figure}

The samples are $p$-doped (100) GaAs quantum well structures symmetrically $p$-doped about a quantum well of 20-nm width. 
The charge carriers are 2D holes of a density of $\sim4\times$10$^{10}$~cm$^{-2}$ and a carrier mobility of $\mu=2.5\times$10$^6$~cm$^2$/(V$\cdot$s). The setup schematically shown in Fig.~\ref{fig:Fig1}(b) includes two separate detections. 
During the measurement, $V_{dc}$ is scanned up to $\pm120$~meV depending on the specific filling factor, with a minimum step size of 10$\mu$V. 
Results of the Hall resistance $\rho_{xy}$ and magnetoresistance $\rho_{xx}$ are shown in Fig.~\ref{fig:appen1}(a). The Corbino geometry provides two choices with the ac lock-in technique: a current ($I_{ac}$) can be applied either from outer leads A-D or inner leads a-d, while the Hall resistance is measured with leads E-B or f-b (or alternatively E-C or e-c). The  $\rho_{xx}$ results measured from the inner and outer edges are identical and consistent with a rectangle Hall bar (topologically equivalent). 

The decoupling of the inner and outer edges is verified by monitoring the changes in the equipotential of the inner edge ($V_{in}$). Fig.~\ref{fig:appen1}(b) shows the transverse voltages drop, $V_{cC}$ and $V_{eE}$, measured individually across the upper and lower arms, in response to an ac current ($I_{ac}\sim1$~nA) running between leads A-D in comparison to the Hall and magnetoresistance. $\rho_{cC}=V_{cC}/I_{ac}$ and $\rho_{Ee}=V_{Er}/I_{ac}$ are approximately even outside the plateau and the variation is achieved through a finite dissipative bulk current.  However, in the central region, $V_{cC}$ and $V_{eE}$ become constant, indicating the absence of any extended bulk current. The inner edge potential can be determined by the difference
\begin{equation}
V_{Cc}-V_{eE}=(V_C-V_{in})-(V_{in}-V_E)=V_C+V_E-2V_{in}    
\end{equation}
Because $V_E=0$ and $V_C=V_H$, $V_{in}$ is written as $[V_H-(V_{Cc}-V_{eE})]/2$. The result confirms that $V_{in}$ is identical to $V_{eE}$, confirming the lower outer edge at the same potential of contact D which is grounded. This is consistent with the Buttiker picture~\cite{Buttiker1988} for highly transparent contacts. 

The constant Hall and magnetoresistance shown in Fig.~\ref{fig:appen1}(b) visually show a different width than that for the constant arm resistance marked with a width $\alpha$. This is simply because small deviations to the Hall resistance are not visible on the shown scales. Fig.~\ref{fig:appen1}(c) shows the $\rho_{xy}$ magnified fifty times in the vertical scale, and the constant $\rho_{xy}=h/e^2$ step shows an identical width with $\alpha$ with the constant $V_{cC}$ and $V_{eE}$ steps. 

\noindent\textbf{B. Transport Electrometer setup for resistance and differential resistance measurement}

Localization is indispensable for stabilizing the QHE against varying $\nu$. Most view the bulk as an enormous insulator offering little features except for energy gaps detectable with spectroscopy techniques. This is only qualitatively true for robust systems under zero or small biases. However, when the system is brought to the verge of a breakdown, proper transport measurement can capture the characteristics of the protection mechanism. The advantage of the technique is that it senses the global effect consisting of random reconstruction processes, which is not accessible with local probe/imaging techniques. For the I-sourcing/V-sensing triaxial (guarded) technique, a Keithley 6430 Femto-Amp ($f$A) is adopted, with an electrometer preamp capable of resolving resistance up to 10~G$\Omega$ with little distortion (owning to a $10^{15}\Omega$ input impedance). Noise filters are also implemented to minimize signal distortion down to 0.05~$f$A (with voltage sensing at 10~$n$V resolutions). This electrometer setup is utilized for the differential resistance $dV/dI$ measurement. As shown in Fig.~\ref{fig:appen1}(e) for $\nu\sim1.003$, though the measured $dV/dI$ detects an insulator with resistivity beyond 50~G$\Omega$ below the breakdown threshold, it captures a remarkable single energy threshold supporting resonant behavior - a single step of 10~$\mu$V increase of $V_{dc}$ around 75 mV results in a plummet over three orders of magnitude in $dV/dI$ from $>$50~G$\Omega$ to 8~M$\Omega$. This behavior provides a key clue for understanding the reconstruction mechanism.

\noindent\textbf{C. Filling factor dependence}

Fig.~\ref{fig:appen1}(d) shows the $V_{th}$ measured for different filling factors measured around 35mK. Note that the filling factors are defined as the B-field normalized by the field corresponding to the center peak. The dashed lines are exponential fittings, slightly asymmetric about $\nu=1.00$, and they extrapolate to zero at $\nu=0.97$ and 1.02. This range of the filling factor corresponds to a B-field from 1.91 and 2.035 Tesla, which is exactly the width $\alpha$ of the Hall plateau shown in Fig.~\ref{fig:appen1}(c). In other words, the non-zero $V_{th}$ also defines the Hall plateau. The breakdown threshold is determined by performing dc-VI or $I/dV$ for a fixed $\nu$ value. Fig.~\ref{fig:appen1}(e) shows the dc-VI curves showing reduced \textbf{E}$_{br}$ for $\nu$ from 1.003 (first on the right) to 1.01 (first on the left). Increasing deviation from integer results in a drastic decrease in \textbf{E}$_{br}$ by approximately 50\% for just a small $\nu$ range. The threshold jumps are higher for $\nu$ closer to the integer, even though the threshold behavior is characteristically unaffected.

\end{document}